# Interface-sensitive nuclear magnetic resonance at a semiconductor heterojunction using hyperpolarization

Atsushi Goto, Kenjiro Hashi, Shinobu Ohki, and Tadashi Shimizu

*National Institute for Materials Science, Tsukuba, Ibaraki 305-0003, Japan*

**Abstract**

We demonstrate an interface-sensitive NMR in a semiconducting nanostructure, where an NMR signal from the minute heterojunction region of a model heterojunction structure ($In_{0.48}Ga_{0.52}P/GaAs$) is detected by using nuclear hyperpolarization created by optical pumping. The key to the detection is the use of minute lattice distortions occurring at the heterojunction due to the lattice mismatch, which enables us to create and localize hyperpolarization at the heterojunction and distinguish it from the other parts. In particular, the suppression of nuclear spin diffusion by the spatial variation in the strain and the resultant unexpectedly stable hyperpolarization at the heterojunction are the keys to successful detection.

The heterojunction is one of the most important features of a semiconductor nanostructure and is the site at which essential functions are carried out in many devices. Nuclear magnetic resonance (NMR), as a powerful analytical tool for materials, is expected to provide essential information on heterojunction properties, but limitations in sensitivity and spatial selectivity have prevented its use in this area. Here, we demonstrate a direct observation of NMR signals from the minute heterojunction region without special treatment of the sample. The scheme utilizes nuclear hyperpolarization created and localized at a part of semiconducting nanostructure [1], which is to be called "hyperpolarization labelling". This study is along the lines of recent efforts for the spatial control of hyperpolarization in optical pumping NMR [2-4].

The implementation of this scheme requires the following: (1) creation, (2) localization, and (3) selective detection of hyperpolarization at a heterojunction. However, all of these steps are challenging mainly because of the low energies (~ MHz, in terms of frequency) and long wavelengths (~ cm) of the radio-frequency fields that resonate with nuclear spins. Recently, Willmering et al. reported the detection of optically pumped NMR signals from the interface between GaAs and aluminum oxide films using quadrupolar satellites induced by strain [4], which shows that strain at a heterojunction provides the key to resolving these problems. Here, we demonstrate the NMR signal detection at a heterojunction between semiconducting layers by hyperpolarization labelling using strain at the junction, mainly focusing on the (1) creation and (2) localization of hyperpolarization.

For demonstration purposes, we used a model heterojunction structure composed of undoped GaAs and $In_{0.48}Ga_{0.52}P$ (InGaP, hereafter) layers, whose lattice constants match at room temperature. Figures 1(a) and 1(b) illustrate the photon absorption and the nuclear spin polarization near the heterojunction, respectively. At sufficiently low temperatures, light with photon energy $E_P$ ~ 1.5 eV and helicity $\sigma^+$ passes through the InGaP layer (band gap ~ 1.9 – 2.0 eV) and is absorbed by the GaAs layer up to a "penetration depth" [5,6], at which hyperpolarization is created. The penetration depth ranges from a few micrometers to hundreds of micrometers depending on the photon energies [7,8], and could be extended by the diffusion of excitons generated in the photon absorption process. The hyperpolarization deep inside the GaAs layer contributes to the central transition of the $^{71}$Ga NMR spectrum due to the cubic symmetry, while that near the heterojunction appears as a quadrupolar-split spectrum due to lattice distortion, which develops upon cooling due to the difference in thermal contraction between the two layers [9,10]. That is, the lattice distortion makes the hyperpolarization at the heterojunction "visible" through the appearance of satellite transitions [4].

In addition, we expect two further effects of the lattice distortion. One is a change in the band-gap energy, which, in particular, lifts the degeneracy at the top of the valence bands (heavy and light holes) as shown in Fig. 1(c). This effect has been described in detail in the case of quantum well structures in Ref. [11]. In general, it introduces complexity in the $E_P$ dependence of the optical pumping NMR spectrum. In some cases, however, it provides selectivity in the optical transitions; e.g., compressive strain renders $E_{HH}$ smaller than $E_{LH}$, which allows us to excite the $E_{HH}$ transition (-3/2 → -1/2) without exciting the $E_{LH}$ transition (-1/2 → 1/2) by tuning the photon energy $E_\text{p}$ to $E_{HH}$. Consequently, the excited electrons predominantly occupy the $m_J = -1/2$ state, resulting in an enhanced hyperpolarization [11]. The other is the formation of a nuclear spin diffusion barrier that reduces leakage of the hyperpolarization. Nuclear spin diffusion is caused by the flip-flop process $(I_+^i I_-^j + I_-^i I_+^j)$ between adjacent nuclear spins [12], which is blocked when the energy for the transition is mismatched between them; this effect is known as the "nuclear spin diffusion barrier" [13-15]. In the present case, the spatially varying electric field gradient (EFG) causes the mismatch [16,17]; the quadrupolar splitting for $I = 3/2$ nuclei is given by $\nu_Q = e^2qQ/2\hbar$, where e$q$ is the principal value of the EFG tensor and $Q$ is the quadrupolar moment. The inhomogeneous distortion provides the variation of e$q$, resulting in the variation of $\nu_Q$. Since the barrier is effective only for satellite transitions, the long relaxation time is essential as well. These effects are expected to work together to reduce the channels for nuclear spin diffusion.

The sample used was fabricated by growing a thin (100 nm) InGaP layer on a semi-insulating undoped GaAs substrate (350 μm) after depositing an undoped GaAs buffer layer (200 nm) using the molecular beam epitaxy method (IntelliEPI, Inc). The NMR measurements were performed at 16 K and 7.05 T using an optical pumping NMR system developed by the authors [18]. The sample setup is illustrated in Fig. 2(a). A piece of the sample (5 mm × 6 mm) was fixed to a sapphire plate on the GaAs side (back side) with grease, mounted on the heat anchor located at the head of the static NMR probe, and set in the vacuum space of the cryostat. The sample was cooled through thermal contact between the heat anchor of the probe and the cold head of the cryostat, which was connected to a GM-cryocooler. At low temperatures, the sample contracted, and the difference in the thermal expansion coefficients (5.9 ×10$^{-6}$ /K for GaAs [9] and 5.3×10$^{-6}$ /K for In$_{0.48}$Ga$_{0.52}$P [10] at 300 K) caused strain at the heterojunction.

The excitation lights were provided by two laser systems: a Ti: sapphire CW tunable laser for the measurement of photon energy dependence, and a diode laser (826 nm) for the relaxation measurements. The linearly polarized light emitted from either laser system was transmitted to the cryostat through polarization-maintaining fibers, and converted to circularly polarized light by a quarter waveplate before illuminating the InGaP surface of the sample in parallel to the magnetic field. The light power was approximately 110 mW at the sample surface.

The pulse sequence used for the relaxation measurements is shown in Fig. 2(b). A set of $\pi/2$-pulses (comb pulse) for the saturation of nuclear magnetization is followed with illumination by the excitation light ($\sigma^+$, 826 nm) for 2 min. The sample is then left in the dark for the duration of $\tau$, during which the nuclear magnetization decays. Finally, a single $\pi/2$ pulse is applied and the free induction decay signal is detected. The $E_p$ dependence of the $^{71}$Ga NMR signal intensity was obtained with a similar sequence, "comb-$\tau_L$-($\pi/2$)-FID" with $\tau_L$= 2 min, but under continuous illumination by $\sigma^+$ light.

Figure 3(a) shows the $E_p$ dependence of the $^{71}$Ga NMR signal intensity ($E_p$ profile) at the main peak. The negative intensity for $E_p$ > 1.490 eV indicates that the polarization is antiparallel to that at thermal equilibrium, as anticipated for $\sigma^+$ light [19,20]. Figure 3(b) shows the $^{71}$Ga spectra measured at four $E_p$ values as indicated by the arrows in Fig. 3(a). Pairs of satellite peaks are seen in the spectra at (i), (ii), and (iii) owing to the first order quadrupolar splitting caused by the lattice distortion. The imbalanced intensities of the two satellites result from low nuclear spin temperatures [4,21]. By contrast, the spectra measured in the dark exhibit no distinct satellites, as shown in Fig. 4(b), indicating that most of the GaAs layer remains close to cubic symmetry. The appearance of the satellites upon illumination indicates that hyperpolarization occurs in the distorted region near the heterojunction. It should be noted that the absolute value of the strain (represented by $\nu_Q$) is not a characteristic property of the sample but depends on the sample setting; the front end of the InGaP layer is a free boundary while the rear end of the GaAs is weakly fixed to the sapphire plate as shown in Fig. 2(a). Consequently, the mismatch between the two layers is resolved by the lattice distortion at the heterojunction.

The relative intensities of the satellites against that of the whole spectrum can be used as a measure of the contribution from the region at the heterojunction [(A): (B) in Fig. 1(b)]; the NMR intensity is roughly proportional to the product of the amplitude of nuclear polarization and the number of nuclei involved. Estimation for spectrum (i) in Fig. 3(b) yields 11.1 % and 10.2% for the lower- and upper-frequency satellites, respectively. Taking into account that the ratio between the central and satellite peaks is 3:4:3, the contribution from the strained region is 37.5% of the total spectral intensity.

Figure 3(c) shows the contour plot of the spectral intensity reconstructed from the spectra measured for Fig. 3(a), which visualizes the relationship between $E_p$ and $\nu_Q$. One finds that the spread of the spectrum due to the satellites gradually changes with $E_p$, which reflects both the distribution of strain near the heterojunction and those of the gap energies caused by the strain as shown in Fig. 1(c).

Let us estimate the distributions of $E_{HH}$ and $E_{LH}$ caused by the strain. The kp-perturbation theory yields the shifts caused by the strain, $\Delta E_{HH} = \delta E_H - \delta E_S/2$ and $\Delta E_{LH} = \delta E_H + \delta E_S/2 - (\delta E_S)^2/(2\Delta_0)$, where $\delta E_H = -9.7\epsilon_\parallel$ (eV) and $\delta E_S = -6.5\epsilon_\parallel$ (eV) are the deformation energy shifts, and $\Delta_0 = 0.34$ eV is the spin-orbit splitting energy ($\epsilon_\parallel$: in-plane strain) [22]. On the other hand, Wood et al. reported the relationship between the $\nu_Q$ value of $^{71}$Ga and the out-of-plane strain ($\epsilon_{zz}$) in GaAs [23]. According to their result, $\nu_Q/2 = 5$–$20$ kHz (the frequency range of the satellites shown in Fig. 3(c)) corresponds to $\epsilon_{zz} = \pm (0.007$–$0.028)$ %, where the sign depends on the strain type (compressive/tensile). Assuming a plane stress with Poisson's ratio: $\nu = 0.32$ [24], it yields $-\Delta E_{HH} = \pm (0.5$–$1.9)$ meV and $-\Delta E_{LH} = \pm (1.0$–$3.8)$ meV. As the shifts of these levels are small, their influence on the $E_p$ profiles is small as well. In fact, the $E_p$ profiles at three frequencies indicated by (I)-(III) in Fig. 3(c) exhibit structures similar to one another as shown in Fig. 3(d). Hence, we attribute the distribution of the spectral intensity in Fig. 3(c) to that of the strain near the heterojunction. Figure 3(e) shows the distribution of strain (in terms of $\nu_Q/2$) deduced from Fig. 3(c). It is found that it spreads out over a wide range of $\nu_Q/2$ with a peak around 9 kHz.

Figure 4(a) shows the result of the relaxation measurements for the $^{71}$Ga spectra measured with the sequence shown in Fig. 2(b) and $E_p$ indicated by the black arrow in Fig. 3(a). The spectra consist of the contributions from the hyperpolarization [(A) + (B) in Fig. 1(b)] and the nuclear spins at thermal equilibrium in the bulk [(C)]. The latter contribution was obtained with the same sequence but without light illumination, and is shown in Fig. 4(b). By subtracting the spectra in Fig. 4(b) from the corresponding ones in 4(a), we can obtain the spectra for the hyperpolarization [(A) + (B)], which is shown in Fig. 4(c).

The most important feature in Fig. 4(c) is the presence of stable satellite peaks; while the absolute intensity of the main peak decays to 1/5 of its initial value in 2 h, the satellites appear unchanged over the same time scale. As a result, the spectral shape approaches a pure quadrupolar-split NMR spectrum at long τ, which represents the hyperpolarization near the heterojunction [(A)].

Bloembergen suggested that the inhomogeneity of the nuclear spin polarization induces the polarization flow to less polarized regions (the bulk GaAs in this case), resulting in rapid nuclear spin relaxation [25]. For a typical diffusion constant, $D = 3 \times 10^{-13}$ cm$^2$/s [26], the hyperpolarization is expected to diffuse up to 100 nm in 5–6 min [15]. In reality, however, the satellites remained unchanged for 2 h. The slow relaxation near the heterojunction implies that nuclear spin diffusion, which rate-limits the process, is ineffective. This conclusion is supported by the observation that the value of $\nu_Q$ remains unchanged; if nuclear spin diffusion to less distorted regions had occurred, it would result in the decrease of $\nu_Q$ at larger τ, and the spectrum would approach a single-peaked shape. Paravastu et al. discovered that, in a bulk GaAs sample with strain near the surface, the asymmetry between the two satellite intensities gradually decreases (the absolute value of the nuclear spin temperature rises) for long light irradiation time (> 30 sec), which can be attributed to the leveling of the high nuclear polarization at the surface by nuclear spin diffusion [21]. Our observation that the asymmetry does not change even after a long period of time in the dark indicates that the nuclear spin diffusion is blocked.

The nuclear spin diffusion barrier is effective only in the region where the EFG gradient is steep. Khutsishvili showed that the diffusion is blocked when the energy mismatch between neighboring nuclei ($\Delta E$) is larger than the dipolar-broadened line width ($\delta\omega_d$) [14]. A rough estimation yields an energy mismatch (in terms of frequency) given by $\Delta E_\mathrm{E} = (\nu_Q/2)(a/\delta L_D)$, with $a$ and $\delta L_D$ being the distance between nuclei and the length over which the distortion is relaxed, respectively, while $\delta\omega_d$ is approximated as $\delta\omega_d \sim \gamma_n^2 \hbar I(I+1) a^{-3}$ [12], in the order of a few kHz [27]. Considering that $\nu_Q/2 \sim 9$ kHz, $\Delta E_{\text{EFG}} > \delta\omega_d$ implies that $\nu_Q$ changes within a few lattice constants.

In conclusion, we have demonstrated that the lattice distortion at the heterojunction provides us with two effects: the visualization of the hyperpolarization through the appearance of satellite peaks and the formation of a diffusion barrier caused by the EFG gradient. Consequently, large hyperpolarization can be localized near the heterojunction, which allows us to perform interface-sensitive NMR.

We would like to thank T. Mano for fruitful discussions. We appreciate the technical assistance by the High Magnetic Field Station (Tsukuba Magnet Laboratory) of NIMS. This work was partially supported by JSPS KAKENHI Grant Number 25287092.

**Figure captions**

FIG. 1. (a) Schematic of the photon absorption at the heterojunction between the undoped GaAs and the InGaP layers. Since the InGaP layer (band gap ~ 1.9 -2.0 eV) is transparent for light with $E_p$ ~ 1.5 eV, light passes through the InGaP layer and reaches the heterojunction, where it is absorbed and hyperpolarization is created. (b) Schematic of the hyperpolarization in the sample. The horizontal and vertical axes correspond to the distance from the surface and the degree of nuclear spin polarization, respectively. Shown in red is the lattice distortion region near the heterojunction, in which quadrupolar broadening occurs. (A)–(C) represent the different regions in GaAs: (A) the distorted and highly hyperpolarized region near the heterojunction, (B) the undistorted and slightly hyperpolarized region and (C) the bulk region at thermal equilibrium. (c) Energy levels at the Γ-point in the presence of compressive and tensile strain and the optical transitions with σ⁺ light. The strain shifts the band-gap energies from that in the absence of strain ($E_G$), and lifts the degeneracy between the heavy- ($m_J = \pm 3/2$) and light-hole bands ($m_J = \pm 1/2$) at the tops of the bands. $E_{HH}$ and $E_{LH}$ are the corresponding band gaps, and the numbers beside the arrows (1, 1/3) are the transition probabilities.

FIG. 2. (a) Schematic of the sample setup. The sample is fixed to a sapphire plate on the back (GaAs) side, and illuminated by circularly polarized light from the front (InGaP) side. Figure not drawn to scale. (b) Pulse sequence for the relaxation measurements.

FIG. 3. (a) Photon energy dependence of the ⁷¹Ga NMR intensity at the main peak ($E_P$ profile) for σ⁺ light. Each point was measured twice (downward and upward $E_P$ sweeps). The solid line is a guide for eyes. The arrows labeled with (i)–(iv) indicate the photon energies at which the spectra in (b) were measured, while the black arrow indicates the photon energy for the relaxation measurements in Fig. 4. (b) The spectra for the photon energies in (a) (upper panel) and its expansion near the baseline to highlight the satellites (lower panel). The horizontal axis represents the deviation from the irradiation

frequency (91.6578 MHz). These spectra were obtained by single shot measurements in the downward $E_p$ sweep and with an exponential apodization in the Fourier transform processing. (c) Contour diagram of the spectral intensity reconstructed from the spectra measured for (a) (downward $E_p$ sweep). The contours around the main peak are omitted (white region) to highlight the satellites. The horizontal slices of the diagram at the $E_p$ values indicated by (i)–(iv) correspond to the spectra in (b), while the vertical slices at the frequencies indicated by (I)-(III) correspond to the $E_p$ profiles in (d). The black arrow indicates the $E_p$ value for the relaxation measurements in Fig. 4. (d) $E_p$ dependences of the spectral intensity at the frequencies indicated by (I)-(III) in (c) normalized at their respective maximum values ($E_p$ profiles). (e) Histogram representing a distribution of strain near the heterojunction deduced from (c). The height at each bin was calculated by integrating the spectral intensity over the $E_p$ range shown by the dashed line in (c) and the bin width of 1 kHz. The height is normalized at the maximum.

FIG. 4. (a) The $^{71}$Ga NMR spectra measured with the sequence shown in Fig. 2 (b) by changing the duration of time in the dark, τ, from 0 min up to 120 min (2 h). The horizontal axis represents the deviation from the irradiation frequency (91.6578 MHz). (b) Spectra measured in the dark (i.e., in the absence of illumination). (c) The spectra for the hyperpolarized nuclear spins obtained by subtracting (b) from (a). The arrows indicate the positions of the satellites. These spectra were obtained by single shot measurements.

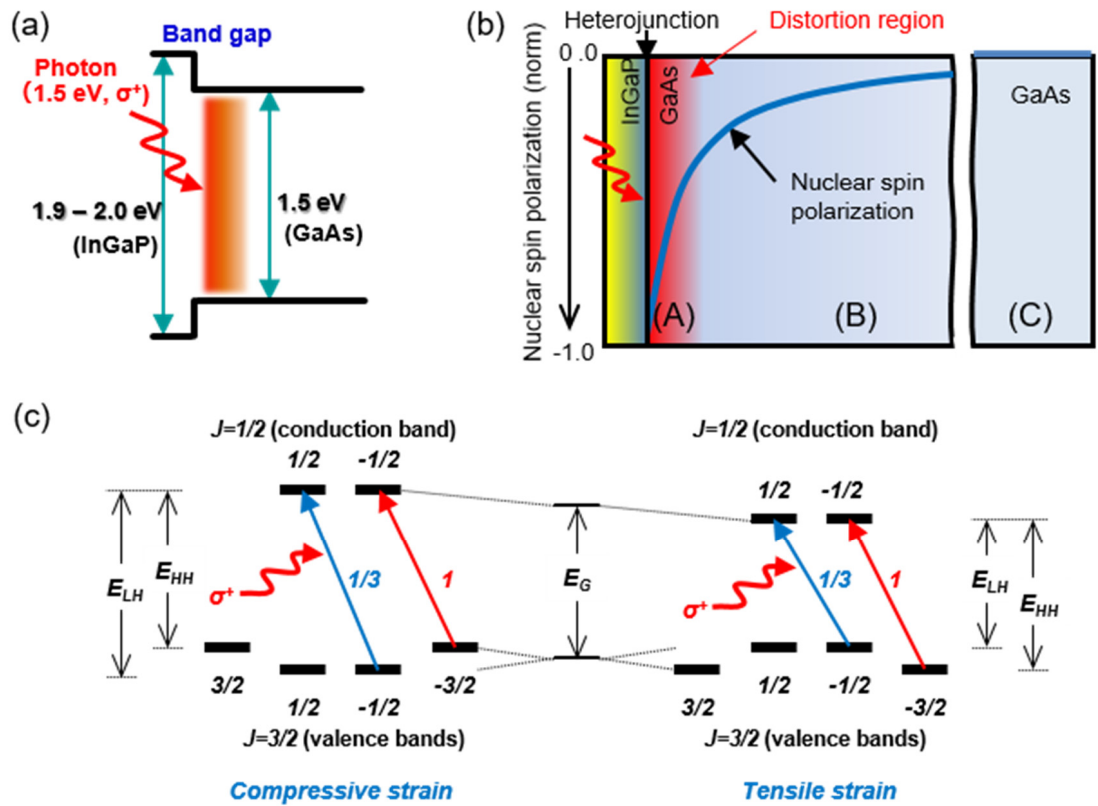

**FIG. 1**

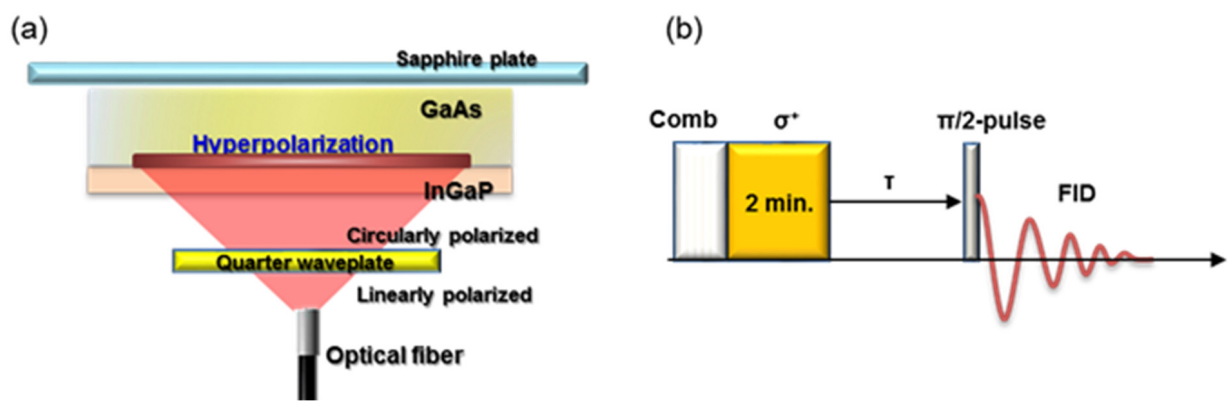

**FIG. 2**

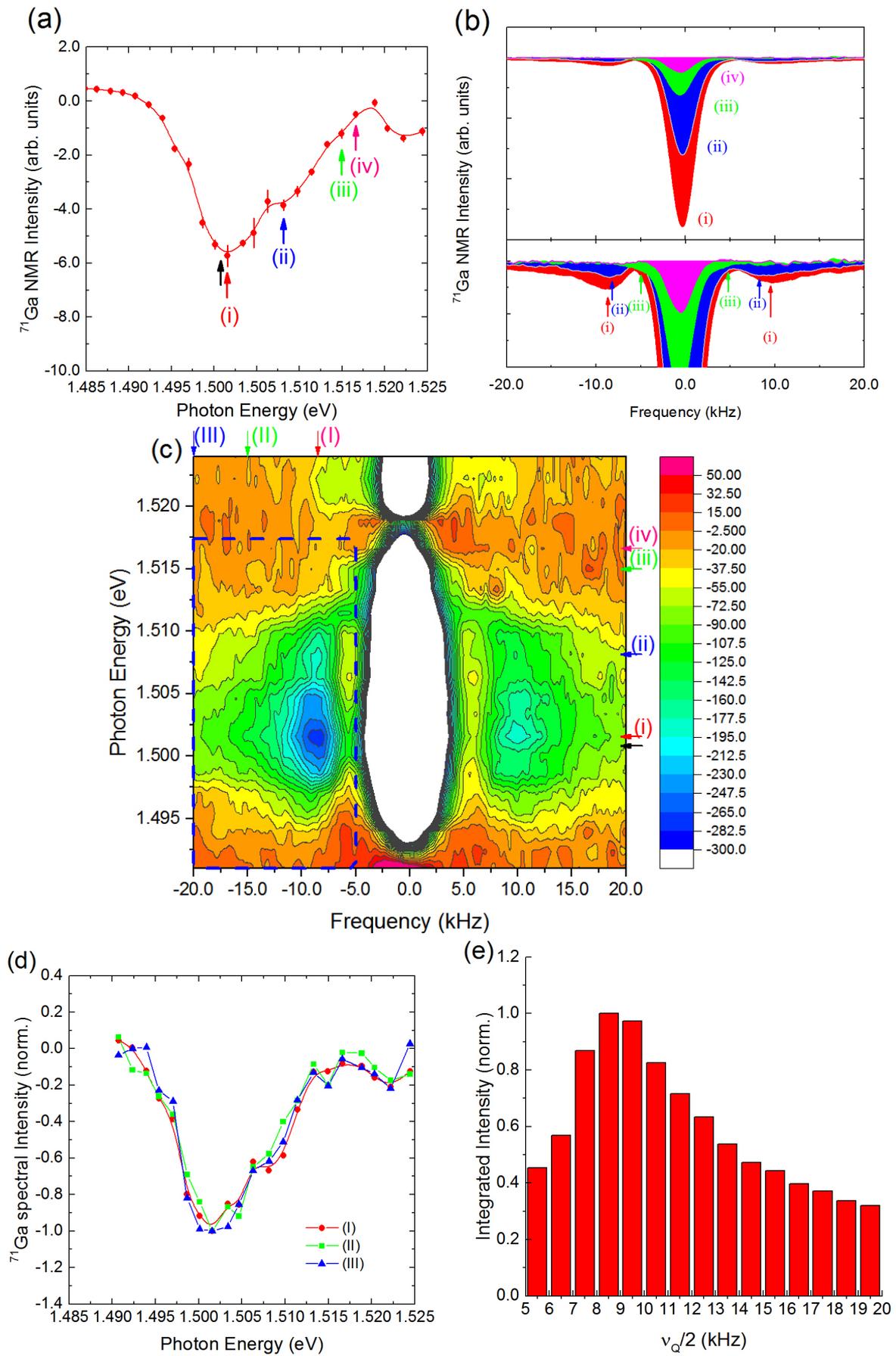

**FIG. 3**

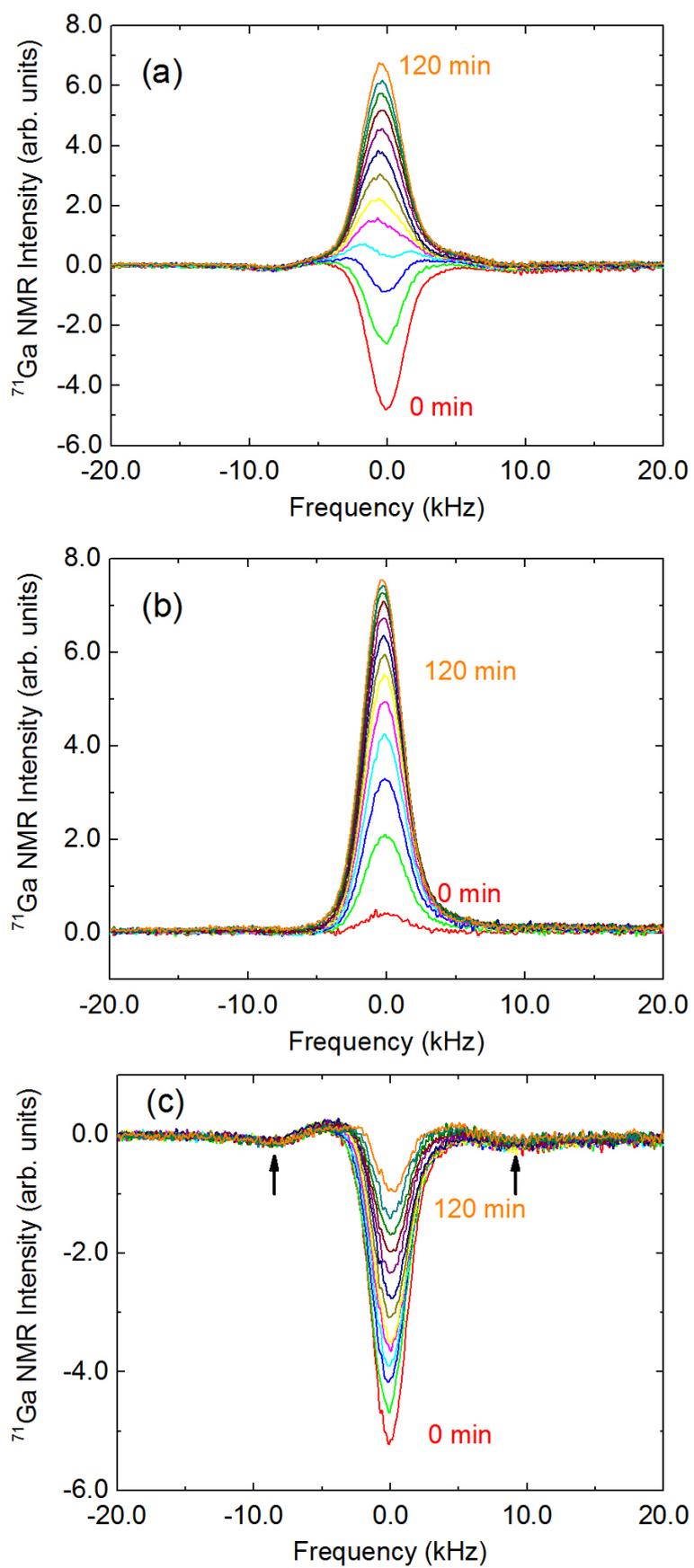

**FIG. 4**